\documentclass[a4paper,fleqn,usenatbib]{mnras}

\usepackage{newtxtext,newtxmath}

\usepackage[T1]{fontenc}
\usepackage{ae,aecompl}

\usepackage{graphicx}	
\usepackage{amsmath}	
\usepackage{amssymb}	
\usepackage{bm}         

\newcommand{\LGCRS}{{\bm l}_\mathrm{GCRS}}
\newcommand{\LLCRS}{{\bm l}_\mathrm{LCRS}}
\newcommand{\LPA}{{\bm l}_\mathrm{PA}}

\title{Earth--Moon VLBI project. Modeling of scientific outcome}

\author[S. Kurdubov et al.]{%
Sergei L. Kurdubov,$^1$\thanks{E-mail: kurdubov@iaaras.ru}
Dmitry A. Pavlov,$^1$
Svetlana M. Mironova,$^1$\newauthor
and Sergey A. Kaplev$^2$
\\
$^1$ Institute of Applied Astronomy of the Russian Academy of Sciences, Saint-Petersburg, Russia\\
$^2$ Federal State Unitary Enterprise ``Central Research Institute for Machine Building'', Korolev, Russia}

\date{Accepted XXX. Received YYY; in original form ZZZ}

\pubyear{2019}

\begin{document}
\label{firstpage}
\pagerange{\pageref{firstpage}--\pageref{lastpage}}
\maketitle

\begin{abstract}
Modern radio astrometry has reached the limit of the resolution that is
determined by the size of the Earth. The only way to overcome that limit is to
create the radio telescopes outside our planet. It is proposed to build an
autonomous remote-controlled radio observatory on the Moon. Working together
with the existing radio telescopes on Earth in the VLBI mode, the new
observatory will form an interferometer baseline up to 410\,000 km, enhancing
the present astrometric and geodetic capabilities of VLBI.
We perform numerical simulations of Earth--Moon VLBI observations operating simultaneously
with the international VLBI network.
It is shown that these observations will significantly improve the precision of determination of Moon's 
orbital motion, libration angles, ICRF, and relativistic parameters. 

\end{abstract}


\section{Introduction and history}

VLBI observations allow to determine various parameters important for astrometry
or geodesy, such as coordinates of extragalactic radio sources, Earth rotation
parameters, and coordinates of stations, with accuracy proportional to $\lambda
/ b$, where $\lambda$ is the wavelength, and $b$ is the baseline between two radio
telescopes. At present, the wavelength is 7.5 mm or more, while the baseline is
bounded by the diameter of the Earth. Expansion towards shorter wavelengths is
hardened due to the atmospheric absorption, frequency standards instability and 
data acquisition system limitations.
Hence the most straightforward way to improve the accuracy is the extension of the baseline outside Earth.
Placing one of the interferometer antennae on the Moon will allow to increase
the baseline by the factor of 60 for the international VLBI network.

The Moon serves as a platform for scientific experiment since Project Apollo.
The most important experiments concerning lunar dynamics and selenodesy include:
lunar laser ranging to five lunar retroreflectors (Apollo 11/14/15, Luna 17/21),
the GRAIL experiment which measured the lunar gravitational field with
unprecedented accuracy, and the Lunar Reconnaissance Orbiter (LRO) spacecraft
that provides high-resolution mapping and altimetry. The lunar lander Chang'e 3,
landed at the end of 2013, is operational to this day, as is LRO. It is evident
that space agencies of different countries are headed towards lunar exploration
and maybe a~habitable lunar baseline in near future. The far side of the Moon is
of particular interest for astrophysical experiments because it is naturally
shielded from Earth's natural and artificial radio frequency interference~\citep{Douglas1985,LRA,SkalskyLRO}.

One of the earliest proposals of a Moon-Earth radio interferometry was made by
\citet{Burns1985,Burns1988}. To this day, no attempt of such an experiment has
been made; however, two orbital VLBI telescopes were built. The first, 8-meter
HALCA~\citep{HALCA}, also known by its program name VSOP, provided VLBI
observations from 1997 to 2003. The second, 10-meter RadioAstron~\citep{RadioAstron},
also known as Spectr-R, launched in 2011 and is still operational. However,
while both HALCA and RadioAstron provided great observations for astrophysics,
neither has been used for the purposes of astrometry.  One reason for this is
the difficulty of precise tracking the spacecraft orbit during the VLBI session,
essential for building the reference frame but not so for obtaining images and
studying the structure of the extragalactic radio sources.

One possible difficulty in VLBI on such a large baseline could be the loss of
correlated flux densities.  However, the Spectr-R results show that at least 160
quasars have very compact structures and high brightness temperatures, enough to
obtain fringes at baselines up to 200\,000 km \citep{Kardashev2017}.

Modern VGOS recording systems can register more than 100x wider bandwith
than Spectr-R formatter (16 MHz) and thus we can expect that the new
instrument will have more than 10x sensitivity and that fringes
will be obtained.

\section{Concept}

\subsection{Pros and cons of lunar environment}
Placing a moving instrument on the lunar surface presents some great challenges,
such as complicated landing and maintenance, temperature jumps, and solar
radiation. On the other hand, in some sense the Moon provides better
experimental conditions than Earth. Having no atmosphere and no oceans, the Moon
has a much more stable rotational motion than the Earth; also, the absence
of atmosphere eases the processing of observations and reduces errors that
come from the ionosphere and troposphere interfering with observed radio
noise. Low lunar gravity will allow lighter construction and machinery.

A spacecraft orbiting the Earth does not suffer from the atmosphere, too;
however, the trajectory of a spacecraft is highly unstable as compared to an
object tied to lunar surface. The biggest source of the instability is the solar
pressure. It does affect the Moon, but its effect to the Moon is much more
subtle, predictable, and stable~\citep{SolarRadiationLunarMotion}.
Furthermore the spacecraft orbit always will be a ``byproduct'' of astrometric observations
whereas precise positioning of lunar based antenna gives significant contribution to selenodesy science.

In order to participate in modern astrometric and geodetic VLBI observations,
an orbital VLBI telescope has to change targets fast. Modern Earth-based VGOS antennas
have slewing speed up to 6 degree per second. Corresponding rotation of spacecraft in space can be achieved
only with the reactive thrusters leading to a very unstable orbit. Also the fuel tanks will limit operational time.

\subsection{Co-location}\label{sec:colocation}

It is recommended that the new radio telescope is co-located with
a~retroreflector, either a panel of corner-cubes~\citep{NPKSPPretroreflectors},
or a next generation single cube retroreflector~\citep{TuryshevRetroreflector,Araki2016}.
Requiring no power on the Moon and no data transmission, the Earth--Moon laser
ranging would help to determine the precise location of the station, and also
independently contribute to study of lunar dynamics, and building of the lunar
reference frame, and testing general
relativity~\citep{MurphyLLR,WilliamsBoggsRatcliff,PavlovWilliamsSuvorkin,INPOP17}.

Adding a GNSS receiver to the new lunar VLBI station will allow to receive
signals from Earth satellites, when the Moon happens to be in the same beam
with Earth when viewed from the satellite. Such observations would allow to
improve the determination of orbits of the GNSS satellites in the celestial
reference frame.

\subsection{Choice of location}\label{sec:location}

Only scientific outcome was considered as the primary criterion for location of
the proposed station on the lunar surface. Other criteria, such as: landing and
deployment issues, local terrain, power supply, proximity to possible lunar
baselines were not considered.

There are multiple reasons for the location to be chosen on the visible side of
the lunar disc. First, it will allow direct Moon-Earth data transfer. Second, it
will allow co-location with a lunar laser ranging (LLR) retroreflector and/or GNSS
receiver (see Section~\ref{sec:colocation}). Third, it will allow a slight
(roughly 9\,000~km) increase in the interferometry baseline, by observing the
radio sources that are visible from the far (as viewed from the Moon) side of
the Earth.

The location on the visible side of the Moon exposes the station to radio
interference from the Earth (natural or artificial). However, the receiving pattern of the lunar 
radio telescope will be narrow, similarly to its Earth
VLBI counterparts, so the radio interference will not be a problem.

Assuming only one instrument of this kind, it should be located close to the
lunar equator so that the radio sources in both southern and nothern hemispheres
are visible.

\begin{figure}
  \includegraphics[width=\columnwidth]{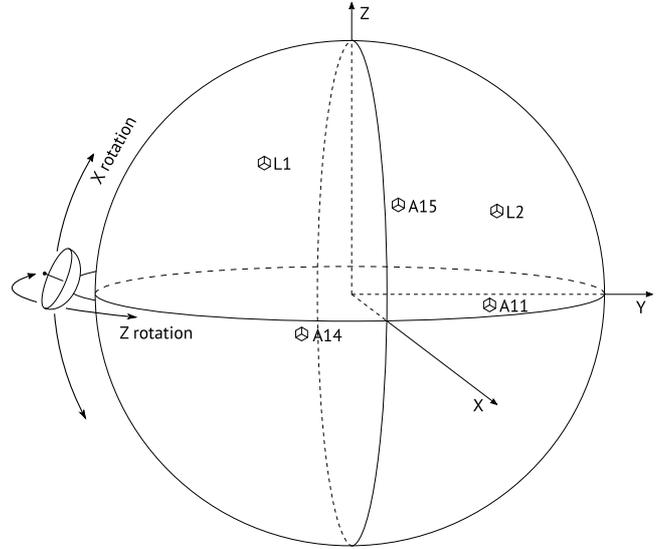}
  \caption{Five existing lunar retroreflector panels and the proposed VLBI
    station. The XY plane and the Z axis are close to lunar equator
    and rotation axis, respectively. The X axis is directed approximately towards the Earth.}
  \label{fig:location}
\end{figure}

The lunar laser ranging---currently the most precise tool to study the lunar
physical librations---has low sensitivity to the rotation of the Moon around the
Moon-Earth direction (see Fig.~\ref{fig:location}). The new VLBI station can
achieve that sensitivity if it is placed outside the central area of the lunar
disc. Another benefit of location outside the central area is that the new
station extends the current lunar reference frame whose accuracy currently
deteriorates outside the central area where are the five retroreflector points.

The chosen location in this work is on the visible part of the Moon's equator,
near its western end. The combination of the existing retroreflectors
and the new VLBI instrument will allow to determine all three ``instant''
(daily/weekly/monthly) rotational corrections to the lunar orientation.
LLR or VLBI alone are able to determine two rotations each.

\section{Decisions on the lunar model}

\subsection{Lunar dynamical model and ephemeris}

We use the dynamical model of the orbital and rotational motion of the Moon
implemented within the EPM planetary-lunar ephemeris
\citep{PavlovWilliamsSuvorkin}. The rotational part of the model, involving
tidal and rotational dissipation, as well as spherical liquid core, was proposed
by \citet{Williams2001}. Later improvements, such as core flattening and
dedicated parameters describing influence of Earth's tides to the orbit of the
Moon, were implemented for DE430 ephemeris~\citep{DE430TR}.  The third publicly
available lunar ephemeris, INPOP, is presently based on the same
model~\citep{INPOP17}.

Lunar ephemerides contain the geocentric (GCRF) position $\mathbf{r}$ and velocity
$\dot{\mathbf{r}}$ of the Moon
as functions of time, and also the Euler angles $\phi$, $\theta$, and $\psi$
and their rates as functions of time.

The lunar frame is aligned with the principal axes (PA) of the undistorted lunar
mantle. The Euler angles define the matrix of the transformation from the lunar
frame to the celestial frame:

\begin{equation}
  R_{\mathrm{L2C}}(t)=R_z(\phi(t))R_x(\theta(t))R_z(\psi(t)).
\end{equation}
$R_x$ and $R_z$ are matrices of right-hand rotations of vectors around axes $x$
and $z$, respectively. The argument $t$ will be omitted when appropriate.

\subsection{Daily lunar orientation parameters}

We study the possibility of determination of daily corrections to the dynamical
model of transformation from the lunar frame to the celestial frame.

\begin{equation}
  R_{\mathrm{L2C}}^{\textrm{daily}}(t)=R_x(\Delta X) R_y(\Delta Y) R_z(\Delta Z) R_{\mathrm{L2C}}(t)
\end{equation}

The daily corrections are assumed to be small and the order of the X, Y, Z
rotations not relevant.

\subsection{Determination of the reference point of the radio telescope}
\label{sec:reference-point-determination}

The VLBI technique requires precise determination of the position of the lunar
radio telescope w.r.t. its Earth counterparts. Such a~determination can be
done in two different ways depending on the task:

\begin{enumerate}
  \item If the task is to build the lunar ephemeris and/or lunar reference
    frame, one should fit to observations the selenocentric position $\LPA$ of
    the lunar radio telescope. The VLBI reduction routine should use the
    determined $\LPA$ to obtain the GCRS position $\LGCRS$:

    \begin{equation}
      \begin{aligned}
        \LGCRS(t) &= \bm r(t) + \LLCRS(t)\left(1-\frac{U(t)}{c^2}\right) \\
                  & - \frac{{\bm{\dot r}}_{\mathrm{B}}(t)\cdot \LLCRS(t)}{2c^2}{\bm{\dot{r}}_{\mathrm{B}}(t)} \\
        \LLCRS(t) &= R_\textrm{L2C}(t)\,\LPA + \bm\Delta(\LPA,t)
      \end{aligned}
      \label{eq:lgcrs}
    \end{equation}
    
\noindent  where $U$ is the gravitational potential at the Moon's center, excluding the Moon's mass,
$\bm{r}_{\mathrm{B}}$ and $\bm{\dot{r}}_{\mathrm{B}}$ are the barycentric position and velocity of the Moon, and
$\bm\Delta$ is the displacement due to solid Moon tide raised by Earth and Sun.

\item For the purposes of astrometry, it is more natural to determine
  geocentric, rather than selenocentric, position of the lunar radio
  telescope. That will give justification to fix and not to determine the lunar
  ephemeris together with the whole Earth--Moon VLBI solution. The link between
  the lunar ephemeris and the geocentric position of the retroreflector is
  relatively small.
  
  The VLBI reduction should then use the determined $\LGCRS(t_0)$ to solve for
  $\LLCRS(t_0)$, then solve for $\LPA$, and finally calculate $\LGCRS(t)$ by
  equation~(\ref{eq:lgcrs}).

  To estimate the accuracy of determination of $\LGCRS(t_0)$, the
  relativistic and tidal terms can be neglected and the derivative matrix of
  $\LGCRS(t)$ w.r.t. $\LGCRS(t_0)$ will be simplified to
  \begin{equation}
    \frac{{\mathrm d} {\bm \LGCRS(t)}}{{\mathrm d} {\bm \LGCRS(t_0)}} \approx R_\textrm{L2C}(t)R_\textrm{L2C}^T(t_0)
    \end{equation}
  
  \end{enumerate}

\section{Setting of experiment}

\subsection{Data}

We simulated two-week intensive Earth--Moon VLBI campaign similar to a subset of
CONT17
(\href{http://ivscc.gsfc.nasa.gov/program/cont17/}{\nolinkurl{ivscc.gsfc.nasa.gov/program/cont17}}).
The ``Legacy-1'' subset was taken; it has 14 VLBI stations in Europe, Russia,
South Africa, Australia, New Zealand, Brazil, Japan, and Hawaii.  To each
scan of Earth radio telescopes observing a source, we add a set of Earth--Moon
VLBI delays of the same source for each of the Earth radio telescopes
participating in the session. Scans involving a quasar not visible from the
lunar radio telescope at the specified time were excluded.  The final dataset 
contains 24\,095 Earth--Moon simulated VLBI delays and 105\,808 real delays on Earth--Earth baselines 
for nine days. The whole campaign lasted 15 days (November 28 -- December 12, 2017), but observations from the remaining six
days of the ``Legacy-1'' subset were not available at the time of writing.

Of the three CONT17 subsets---``Legacy-1'', ``Legacy-2'', and
``VGOS-Demo''---the first one has the best expected accuracy on the EOP determination.  For
comparison, we also took the ``Legacy-2'' subset in the simulation of the
determination of lunar orientation parameters and PPN parameter $\gamma$
(see sections~\ref{sec:estimating-lop} and~\ref{sec:ppn}).
36\,589 Earth--Moon delays were generated and used for that subset.

Simulations of Earth--Moon VLBI delays for different subsets were done
separately, not simultaneously, due to the time overlap.

For one part of the experiment (see~Section~\ref{sec:estimating-ephemeris}), we
used the real LLR data spanning from 1970 to the end of 2017. The most important
observations at present are performed at Apache Point Observatory
\citep{murphy12,MurphyLLR} and Observatoire de la C\^ote d'Azur \citep{OCALLR,
  GrasseLLRIR}. More information about re-weighting and reductions of LLR
observations is given in \citep{PavlovWilliamsSuvorkin}. Some observations were
considered erroneous and were filtered out. Most often the observations from the
McDonald Laser Ranging Station (1988--2015) did not fit well into the model. Of
total 25\,535 LLR observations, 773 were filtered out and 24\,762 were used in the
lunar solution.

\subsection{Software}

Two independent software packages were used. ERA-8~\citep{ERA8} was used to fit
the parameters of the lunar model to observations, and to integrate lunar
ephemeris~(Section~\ref{sec:estimating-ephemeris}). An extension was made to
ERA-8 to process Earth--Moon VLBI delays (in addition to LLR normal points) as
part of a global lunar solution.  QUASAR~\citep{QUASAR} was used to obtain the
VLBI solution for celestial reference frame, using a fixed
lunar ephemeris. An extension was made to QUASAR to bring a VLBI station to the
Moon, determine its location in Section~\ref{sec:estimating-refsyst} or
the lunar orientation parameters in Section~\ref{sec:estimating-lop}.

\subsection{Estimating the improvement of the lunar ephemeris}
\label{sec:estimating-ephemeris}

Two estimations of the accuracy of the lunar ephemeris were obtained: the first
using real LLR observations (1970--2017), and the second using LLR and 15 days of
simulated VLBI observations in the end of 2017.

The celestial and terrestrial VLBI frames were fixed. The formal error of the
simulated Earth-VLBI observations was set to 7 mm ($1\sigma$), similarly to the
postfit RMS discrepancy of the ionosphere-free combination for global VLBI solution.

Each of the determined parameters falls into one of the two categories:
\begin{enumerate}
  \item Dynamical parameters: initial Euler angles of lunar physical libration
    and their rates; initial GCRS position and velocity of the Moon; initial
    core angular velocity; undistorted $J_2$, $C_{32}$, $S_{32}$, $S_{33}$ of the
    Moon; ratios between undistorted main moments of inertia of the Moon; lunar
    core flattening and friction coefficient; two Earth tidal delays; GM of the
    Earth--Moon system.
  \item Reduction parameters: positions and velocities of LLR stations;
    selenocentric (PA) positions of the lunar retroreflectors; Love number $h_2$
    of the Moon; 28 specific biases for different time intervals. The
    selenocentric lunar radio telescope position was determined as well in the
    solution with 15 days of simulated VLBI observations.
  \end{enumerate}

The details about the determined parameters are given in
\citep{PavlovWilliamsSuvorkin}, however, for this work, two changes were
made. One of them is the aforementioned determination of $J_2$. Not fixing
$J_2$ to a GRAIL-determined value brings more realistic uncertainties into the
lunar solution. The second change is that the additional three kinematic terms
of the longitude libration (with amplitudes in several mas) were not determined
and thus absorbed into other dynamical parameters' formal errors.

Parameters from both categories were determined simultaneously using the
least-squares method, their formal errors and covariance matrix were calculated.

In addition to the dynamical parameters, the dynamical model also has constant
undistorted Stokes coefficients of Earth's gravitational potential (up to degree
6) taken from EGM2008~\citep{EGM2008} solution, and Moon's gravitational
potential ($C_{30}$, $C_{31}$, $S_{31}$, and others of degrees 4--6) from
GL1200~\citep{GRGM1200a} solution. From the latter solution, the lunar $k_2$ Love
number is also used. 

The covariance matrix of the parameters under variation was formed as
a submatrix of the lunar solution covariance matrix with excluded reduction
parameters, to which rows and columns were added corresponding
to the constants. The resulting matrix $\mathbf{M}$ had size $(n+k)\times (n+k)$,
where $n = 28$ was the number of the dynamical determined parameters
and $k = 83$ was the number of constants. 

In the added rows and columns, squares of the formal errors
of corresponding constants were assigned to the diagonal terms,
while the off-diagonal terms were set to zero.

The Monte-Carlo simulation of the lunar ephemeris was done in the following
way. The Cholesky decomposition of the covariance matrix was calculated:
$\mathbf{M} = \mathbf{L}^\mathrm{T}\mathbf{L}$, where $\mathbf{L}$ is a lower
triangular matrix. Each sample $X$ for the simulation was obtained as a sample
from the multidimensional normal distribution described by $\mathbf{M}$: $X =
\mathbf{L}^\mathrm{T} Y$, where $Y$ is a $(n + k)$-vector sample of independent
normally distributed random variables with zero mean and unit variance.

For each of the sampled $X$, a corresponding ephemeris of orbital and rotational
motion of the Moon was obtained by numerical integration. The integrated
dynamical model contained not only the Moon, but also the Sun, all planets, Pluto,
Ceres, Pallas, Vesta, Iris, and Bamberga. The parameters related to non-lunar
part of the model were not varied, because their influence on the lunar ephemeris is
relatively small, and their influence on the lunar ephemeris uncertainty
is negligibly small.

The chosen epoch was January 2, 2018, and numerical integration was for five
years forward (2018--2022). The scatter of the determined initial parameters of
the Moon (position and Euler angles) and the evolution of those parameters over
time was studied. Two different Monte-Carlo simulations were done, one with only
LLR observations, and the other with LLR and simulated VLBI observations.
1\,000 samples were generated for each of the two simulations.

\subsection{Estimating the improvement of the celestial reference frames}
\label{sec:estimating-refsyst}

In order to estimate the accuracy of celestial reference frame we perform
the solution over all CONT17 data with source right ascension and declinations as global parameters.
The tropospheric delay, clock synchronisation parameters, LOP and EOP were estimated for each session.

For the Earth baselines we also use the model noise instead of real VLBI observations because
the real VLBI data depends heavily on clock corrections
and the procedure of determination of those corrections should work
with consistent O-C values. Since do not have Earth--Moon VLBI data
consistent with real Earth VLBI clock, we took the correlator estimated
noise for all observations. The noise was multiplied by the factor of 4,
which is an ad-hoc value that represents the clock, troposphere etc. scatter
from real world and was found from comparison of parameter errors
obtained in Earth VLBI processing with model noise and real VLBI data.

\subsection{Estimating the determination of the lunar orientation parameters}
\label{sec:estimating-lop}

LOP estimation was done in a similar manner to the celestial reference frame
determination (sec.~\ref{sec:estimating-refsyst}).

With the chosen location of the lunar radio telescope
(Section~\ref{sec:location}), its position in the celestial reference frame will
be sensitive to $\Delta X$ and $\Delta Z$ daily corrections but not to
$\Delta Y$.

The $\Delta X$ and $\Delta Z$ corrections were assumed independent and constant
for each of the nine days of the ``Legacy-1'' subset of the CONT17 campaign with
simulated Earth--Moon VLBI data.

For comparison, similar simulation was done with the ``Legacy-2'' subset
spanning 15 consecutive days.

Lunar laser ranging can also independently contribute to determination of the
lunar orientation parameters. However, Earth--Moon ranges are sensitive to Y and
Z rotations but have almost no sensitivity to X rotations. As shown
by~\citet{PavlovYagudina}, the formal errors of daily (nightly) $\Delta Y$ and
$\Delta Z$ obtained on real LLR data are 1--2 mas at best.

\section{Results of parameter estimation}

\subsection{Lunar ephemeris, lunar reference frame, and Moon-Earth reference frame}

\begin{table}
	\centering
	\caption{Uncertainty of the selenocentric reference points in two solutions. A = Apollo,
          L = Lunokhod, LRT = lunar radio telescope}
	\label{tab:lunar-points-selenocentric}
	\begin{tabular}{lrrr} 
		\hline
		Coordinate & $3\sigma$ (LLR only) & $3\sigma$ (LLR and VLBI) \\
		\hline
                A11 X & 14.5 cm & 0.5 cm \\
                A11 Y & 19.4 cm & 2.0 cm \\
                A11 Z &  5.3 cm & 1.4 cm \\
                L1 X  & 15.7 cm & 0.5 cm \\
                L1 Y  & 13.9 cm & 2.3 cm \\
                L1 Z  &  8.5 cm & 4.8 cm \\
                A14 X & 12.8 cm & 0.5 cm \\
                A14 Y & 19.9 cm & 2.0 cm \\
                A14 Z &  5.4 cm & 1.4 cm \\
                A15 X & 11.5 cm & 0.4 cm \\
                A15 Y & 18.8 cm & 1.6 cm \\
                A15 Z &  7.9 cm & 3.5 cm \\
                L2 X  & 14.0 cm & 0.4 cm \\
                L2 Y  & 16.5 cm & 2.2 cm \\
                L2 Z  &  7.5 cm & 3.5 cm \\
                LRT X & N/A     & 1.5 mm \\
                LRT Y & N/A     & 0.9 mm \\
                LRT Z & N/A     & 0.6 mm \\
		\hline
	\end{tabular}
\end{table}
\begin{table}
	\centering
	\caption{Uncertainty of the geocentric (at epoch 02.01.2018) reference points in two solutions. A = Apollo,
          L = Lunokhod, LRT = lunar radio telescope}
	\label{tab:lunar-points-geocentric}
	\begin{tabular}{lrrr} 
		\hline
		Coordinate & $3\sigma$ (LLR only) & $3\sigma$ (LLR and VLBI) \\
		\hline
                A11 X &  3.9 cm & 2.0 cm \\
                A11 Y &  2.9 cm & 0.9 cm \\
                A11 Z & 14.0 cm & 1.3 cm \\
                L1 X  &  3.1 cm & 2.2 cm \\
                L1 Y  &  3.5 cm & 1.7 cm \\
                L1 Z  & 14.1 cm & 4.4 cm \\
                A14 X &  3.1 cm & 1.9 cm \\
                A14 Y &  3.2 cm & 1.1 cm \\
                A14 Z & 14.4 cm & 1.4 cm \\
                A15 X &  2.6 cm & 1.4 cm \\
                A15 Y &  3.8 cm & 1.7 cm \\
                A15 Z & 13.6 cm & 3.2 cm \\
                L2 X  &  3.7 cm & 2.2 cm \\
                L2 Y  &  3.2 cm & 1.4 cm \\
                L2 Z  & 13.7 cm & 3.2 cm \\
                LRT X &   N/A   & 0.9 mm \\
                LRT Y &   N/A   & 1.7 mm \\
                LRT Z &   N/A   & 1.0 mm \\
		\hline
	\end{tabular}
\end{table}

Positions of lunar reference points were determined simultaneously
with parameters of orbit and physical libration of the Moon, and
other parameters (see~Section~\ref{sec:estimating-ephemeris}).

Table~\ref{tab:lunar-points-selenocentric} shows the accuracy of the lunar
reference frame implemented by five retroreflector points (in case of real LLR
data) or five retroreflector plus one radio telescope point (with real LLR and
simulated VLBI data). One can see a 6x--10x improvement in accuracy of the
existing five points from the VLBI data. One of the key factors of such a major
improvement is that the VLBI observations are ``omnidirectional'', i.e. they
measure the projections of the ``Earth observatory -- Lunar radio telescope''
vector to all directions, and so they are sensitive to the orbital position of
the Moon.  LLR observations are sensitive to the Earth--Moon distance but not
much to the position of the Moon on orbit.

Table~\ref{tab:lunar-points-geocentric} is similar to
Table~\ref{tab:lunar-points-selenocentric} but it shows the accuracy in the
geocentric at epoch (see Section~\ref{sec:reference-point-determination}).  In
the LLR-only solution, the geocentric positions of retroreflector are determined
$\approx$ 1.6x better than the selenocentric ones. It is known that the X coordinate of each
lunar retroreflector panel strongly correlates with the GM of the Earth--Moon
system and also with the semimajor axis of the Moon at the epoch \citep[see
  e.g.][]{DE430MoonTR}. The correlation causes the selenocentric position of a
retroreflector at epoch \textit{and} the position of the Moon at epoch to be
detected with less accuracy than the geocentric position of the retroreflector
at epoch.

As for the ``LLR+VLBI'' solution, the geocentric and selenocentric positions
of the six lunar points are determined with similar accuracy overall, thanks
again to the VLBI measuring in all directions.

\begin{figure}
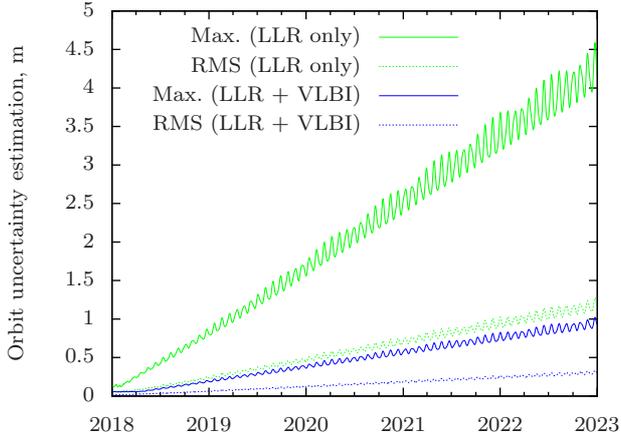

  \input uncertainty-orbit.tex
  \caption{Estimation (maximum and RMS) of the uncertainty of the orbital position of the Moon
    for five years, assuming no observations since 2018, using only LLR data or LLR data with
    simulated VLBI data}
  \label{fig:uncertainty-orbit}
\end{figure}

\begin{figure}
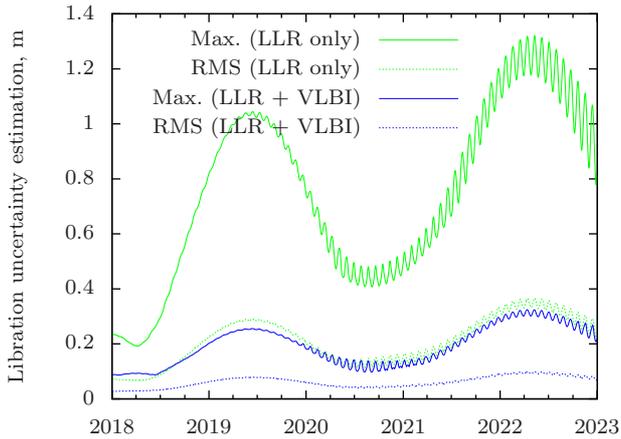

  \input uncertainty-libration.tex
  \caption{Estimation (maximum and RMS) of the uncertainty of the lunar physical libration
    (measured in maximum discrepancy across the lunar surface)
    for five years, assuming no observations since 2018, using only LLR data or LLR data and
    simulated VLBI data}
  \label{fig:uncertainty-libration}
\end{figure}

To estimate the accuracy of either lunar or Moon-Earth reference frame not at
epoch, but for some time into the future, one can look at
Figs.~\ref{fig:uncertainty-orbit} and~\ref{fig:uncertainty-libration}. Maximum
and root-mean-square deviations of sampled ephemeris (see
Section~\ref{sec:estimating-ephemeris}) w.r.t. the nominal ephemeris are
plotted. In five years without new observations, the accuracy of the Moon's
orbital position determined via 47 years or LLR degrades to 4.6~m (maximum);
when LLR is combined with nine days of Earth--Moon VLBI observations, the maximum
error drops to 0.6~m. Similarly, the 5-year maximum error of lunar physical
libration is 1.3~m with LLR and 0.15~m with LLR+VLBI.

\subsection{Lunar orientation parameters}

\begin{figure}
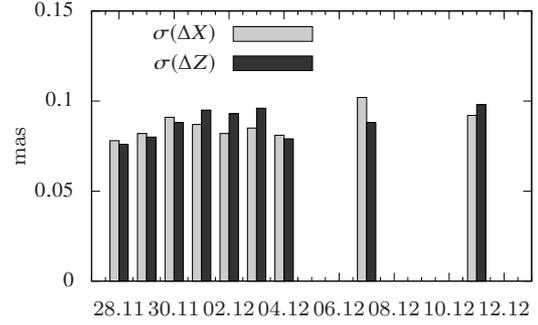

  \input lop-b.tex
  \caption{Formal errors ($1\sigma$) of determined lunar orientation
    parameters (X and Z rotations) in ``CONT17 Legacy-1 with lunar radio telescope''
    scenario}
  \label{fig:lop-b}
\end{figure}

\begin{figure}
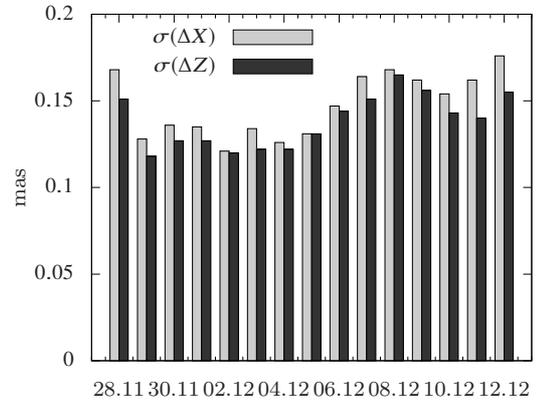

  \input lop-a.tex
  \caption{Formal errors ($1\sigma$) of determined lunar orientation
    parameters (X and Z rotations) in ``CONT17 Legacy-2 with lunar radio telescope''
    scenario}
  \label{fig:lop-a}
\end{figure}

The formal errors of the daily $\Delta X$ and $\Delta Z$ detected from
Earth--Moon VLBI simulations are shown at Figs.~\ref{fig:lop-b} (``Legacy-1''
subset of CONT17) and~\ref{fig:lop-a} (``Legacy-2''). It can be seen that both
rotations are determined with uncertainty (3x the formal error) 0.3 mas or
lower with ``Legacy-1''. With the other subset, the uncertainties are $\approx$
1.5x bigger, despite the larger number of observations than in the first subset.
This comparison agrees with the original CONT17 estimations for the EOP
determination.

\subsection{Celestial reference frame}

\begin{figure}
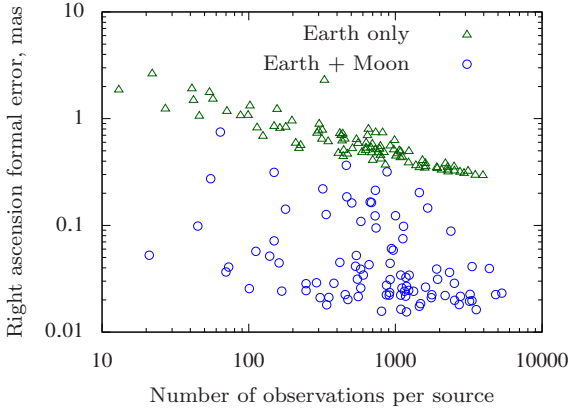

  \input crf-ra.tex
  \caption{Formal errors ($1\sigma$) of determined right ascensions
    of observed radio sources from CONT ``Legacy-1'' observations,
    with and without Earth--Moon VLBI}
  \label{fig:crf-ra}
\end{figure}

\begin{figure}
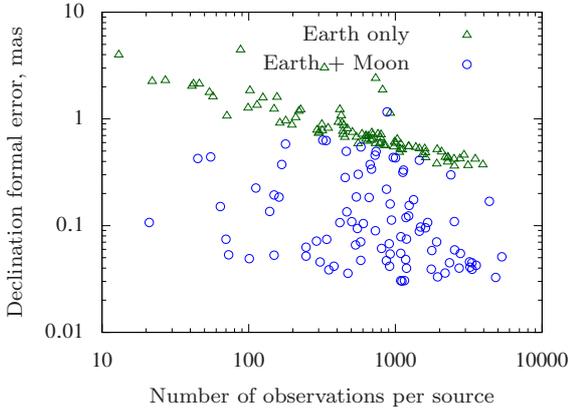

  \input crf-de.tex
  \caption{Formal errors ($1\sigma$) of determined declinations
    of observed radio sources from CONT ``Legacy-1'' observations,
    with and without Earth--Moon VLBI}
  \label{fig:crf-de}
\end{figure}

In order to show the impact of lunar VLBI observation on improvement of celestial reference frame realization
we estimate radio source positions using only CONT17 Earth based observations and observations with Earth--Moon baselines.
Results are presented on Figs.~\ref{fig:crf-ra} and~\ref{fig:crf-de} for right ascension and declination, respectively.
The plots show formal errors of source positions for Earth and Earth--Moon observations.
One can see that for moderate number of sources one Moon-based telescope improves positions accuracy by more than ten times.

The formal errors for the best sources of the ICRF are about tens of
microarcseconds \citep{ICRF2}, which is smaller than the ``Earth-only'' results
on Figs.~\ref{fig:crf-ra} and~\ref{fig:crf-de}.  That is because our estimates
are based on nine days' data, while the ICRF is built using the results from 40
years of VLBI observations.  The Earth--Moon VLBI results, also estimated from
nine days of observations, will improve with longer timespan.

\begin{figure}
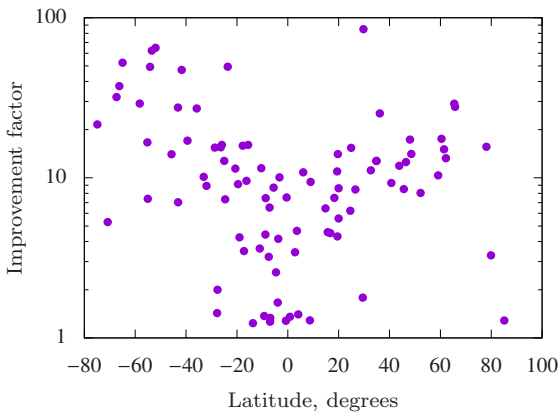

  \input accuracy-de-lat.tex
  \caption{Improvement factor (Earth and Earth--Moon VLBI vs just Earth VLBI) of radio source declination formal error normalized by square root of relative number of observations (as a function of ecliptic latitude).}
  \label{fig:accuracy-de-lat}
\end{figure}

Fig.~\ref{fig:accuracy-de-lat} shows the ratio of the formal errors of determined declinations 
without Earth--Moon VLBI to those with Earth--Moon VLBI, relative to ecliptic latitude.  
The ratio were multiplied by square root of number of observations with Earth--Moon VLBI divided by 
number of observations without Earth--Moon VLBI. 
One can see that points on Fig.~\ref{fig:accuracy-de-lat} are greater than 1 for all latitudes and more than 10 for about half 
of the sources. 
The improvement are the least for latitudes close to zero
because the Earth--Moon baseline and source unit vectors are both close to the ecliptic plane.

Interstellar scattering and intrinsic source structure effects may limit
position errors obtained by Earth--Moon VLBI to greater than those shown in
Figs. \ref{fig:crf-ra} and \ref{fig:crf-de}. However, the Earth--Moon VLBI can
help to estimate the source structure, in the manner similar to present
VLBI observations \citep[see e.g.][]{Imaging}.  There are techniques in
development to apply the source structure map to astrometric VLBI delay
calculation.

\subsection{General relativity tests}\label{sec:ppn}
Relativistic contribution by Solar system bodies to VLBI time delay arises with baseline $b$. 
Using the equation from \citep{1991Klioner}, we computed the gravitational delay and delay due to translational motion
for baselines $b = 6\,000$~km and $b = 380\,000$~km.
Table \ref{tab:delay-g-m} shows the delays for different baseline lengths (the second column) 
different angles between source unit vector and baseline vector (the third column). 
It is visible that delay generated by the Sun is detectable for two antennas located on Earth only if the angle is 
less than 30 degrees. Locating one of the antennas on the Moon will make that delay detectable on all angles. 

\begin{table}
	\centering
	\caption{Delay generated by gravity of a body ($\Delta t_G$) and by its translational motion ($\Delta t_M$)}
	\label{tab:delay-g-m}
	\begin{tabular}{lrrrr} 
		\hline
		Body & Baseline, km & Angle & $\Delta t_G$ & $\Delta t_M$ \\
		\hline
        Sun &   6\,000 & grazing ray &   169.24 ns & 0.0082 ps \\
        Sun &   6\,000 &       1 deg &    45.33 ns &           \\
        Sun &   6\,000 &      30 deg &     1.47 ns &           \\
        Sun &   6\,000 &      90 deg &     0.40 ns &           \\
        Sun &   6\,000 &     175 deg &    17.25 ps &           \\
        Sun & 380\,000 & grazing ray &  8\,588.24 ns & 0.4166 ps \\
        Sun & 380\,000 &       1 deg &  3\,098.72 ns &           \\
        Sun & 380\,000 &      30 deg &    93.59 ns &           \\
        Sun & 380\,000 &      90 deg &    25.02 ns &           \\
        Sun & 380\,000 &     175 deg &     1.11 ns &           \\
		\hline
	\end{tabular}
\end{table}

The model of VLBI time delay contains PPN parameter $\gamma$ which characterize space curvature produced by unit rest mass.
The accuracy of $\gamma$ estimated by initial CONT17 ``Legacy-1'' VLBI series $2.0 \cdot 10^{-3}$ has been
obtained, while the estimation from ``Legacy-1''  and ``Legacy-2''  gives $1.3 \cdot 10^{-3}$. 
Adding Earth--Moon VLBI improves the accuracy to $0.5 \cdot 10^{-3}$ for ``Legacy 1'' and to $0.06 \cdot 10^{-3}$ for
``Legacy-1''  and ``Legacy-2''.

The experiment was not specially designed for estimation of $\gamma$ and does not have sources close to the Sun or Jupiter.
Nevertheless, we obtain improvement factor from 4 to 21. A dedicated experiment like \citep{titov2018}
with Earth--Moon VLBI can overcome the presently best accuracy $0.23 \cdot 10^{-4}$ obtained from 
frequency shift measurements of Cassini spacecraft \citep{Bertotti2003}.

\cite{1980Epstein} also show that post-post-Newtonian deflection of the Sun is about 11 $\mu$arcsec 
which corresponds to delay equal to 1.1 ps for $b = 6\,000$ km 
and 67.6 ps for $b = 380\,000$ km. 
Therefore growth of the baseline will increase the accuracy of relativistic effects detection.

\section{Circumlunar orbit determination simulation}

A dedicated numerical simulation was performed to study the impact of lunar
reference frame improvement (Table~\ref{tab:lunar-points-selenocentric}) to the
determination of orbits of an example lunar satellite constellation. Nine
satellites were modeled with circular inclined orbits with semimajor axis of 4500~km.
The model observations were: inter-satellite links (ISL) measurements and two-way range
measurements to the retroreflector points.

Table \ref{tab:range-accuracy} shows the accuracy of given constellation orbit determination projected
to the average lunar surface point line-of-sight for two scenarios of reference
points uncertainties according to Table~\ref{tab:lunar-points-selenocentric}
columns. ISL measurement accuracies were assumed to be at a few centimeter level
($1\sigma$) to make the results sensitive to the differences between the reference points
uncertainties. Solution is produced with the least squares method using the
measurement interval of 3.5 hours equals to about the half of orbital period and
the following propagation interval of 12 hours. Statistics was taken over all
satellites in the constellation. Three values are provided for each time
interval: root-mean-square value, maximum value over 95\% confidence interval, and
maximum value overall.

\begin{table}
  \centering
  \caption{Range accuracy of orbit determination example based on ISL and two-way range measurements to the reference points with different uncertainties}
  \label{tab:range-accuracy}
  \begin{tabular}{lrr} 
    \hline
Reference points by:  & LLR only & LLR and VLBI \\
    \hline
    RMS (3.5 h measurements) & 7.9 cm & 2.7 cm \\
    95\% CI (3.5 h measurements) & 12.9 cm & 4.8 cm \\
    Max. (3.5 h measurements) & 16.2 cm & 8.0 cm \\
    RMS (12 h propagation) & 14.8 cm & 7.8 cm \\
    95\% CI (12 h propagation) & 26.5 cm & 14.5 cm \\
    Max. (12 h propagation) & 43.3 cm & 24.4 cm \\
    \hline
  \end{tabular}
\end{table}

The results show a $\approx$ 2x improvement in orbit determination accuracy
after suggested Earth--Moon VLBI implementation. 

\section{Conclusion}

For the purposes of astrometry, the placement of the lunar radio telescope near
equator, close to the edge of the visible side of the lunar disc is preferred.

The obtained results are based on the assumption that the sensitivity
of the proposed Earth--Moon interferometer will be enough for obtaining accurate time delay
(i. e. the existence of sufficiently bright and compact quasar structures is assumed).

Just nine days of intensive Earth--Moon VLBI observations will improve the
accuracy of the Earth--Moon reference frame 3x--6x, and the accuracy of the lunar
reference frame 4x--10x. The accuracy of the lunar ephemeris (both orbit and
physical libration), presently calculated from 47 years of LRR, will improve from meters to
decimeters for five-year interval into the future.

The accuracy of daily corrections to lunar orientation reach 0.3 mas ($\approx$
2.5 mm on the lunar surface) for $X$ and $Z$ rotations. With the chosen
location of the instrument, $Y$ rotations can not be determined from Earth--Moon
VLBI; however, LLR is capable of determining them, while with accuracy of 3--5 mas
(centimeters on the lunar surface).

According to simulated observations for nine days, 
the accuracy of the celestial reference frame can be improved more than 10x for about
half of the radio sources distant from ecliptic. The Earth--Moon interferometer will have higher sensitivity to 
the relativistic parameters of post-Newtonian and post-post-Newtonian formalism and can be used for relativistic tests.

The lunar reference frame, improved by the Earth--Moon VLBI, will bring a
$\approx$ 2x improvement in accuracy of the determination of orbits of a
possible lunar satellite constellation. However, such an improvement, while
important for fundamental lunar science, will not be critical for applications
like lunar navigation with the present requirements.

Some of the results promised by Earth--Moon VLBI relevant to study of the Moon
itself could in principle be accomplished with other tools (more precise LLR,
lunar radio transponders, lunar optical telescope), while orbital radio telescopes
are useful for astrophysics. However, the Earth--Moon VLBI will stand unrivaled
for future radio astrometry and building of the Earth--Moon reference frame.

\section*{Acknowledgements}

Authors from the IAA RAS are thankful to Iskander Gayazov, Alexander Ipatov,
Dmitry Vavilov, and numerous other colleagues for helpful discussion, advice,
and support throughout this work.

The current state of lunar study, used in this work as a starting point, would
not have been possible without the effort of personnel at observatories doing
lunar laser ranging. Similarly, the VLBI data used in this work was a result of
a many-year effort of different people and organizations from different
countries. Particularly, the Continuous VLBI Campaign managed by the IVS
provides an inspiring insight of capabilities of international collaboration.


\bibliographystyle{mnras}
\bibliography{moon_vlbi}



\bsp	
\label{lastpage}
\end{document}